\documentclass[12pt]{iopart}
\usepackage{iopams}

\newtheorem{thm}{Theorem}
\newtheorem{lem}{Lemma}

\begin{document}

\title{Optimal dense coding with mixed state entanglement}

\author{Tohya Hiroshima
\footnote[1]{tohya@frl.cl.nec.co.jp}
}

\address{Fundamental Research Laboratories,\\ 
NEC Corporation,\\
34 Miyukigaoka, Tsukuba 305-8501, Japan}

\begin{abstract}
I investigate dense coding with a general mixed state on the Hilbert space $%
C^{d}\otimes C^{d}$ shared between a sender and receiver. The following
result is proved. When the sender prepares the signal states by mutually
orthogonal unitary transformations with equal {\it a priori} probabilities,
the capacity of dense coding is maximized. It is also proved that the
optimal capacity of dense coding $\chi ^{*}$ satisfies $E_{R}(\rho )\leq
\chi ^{*}\leq E_{R}(\rho )+\log _{2}d$, where $E_{R}(\rho )$ is the relative
entropy of entanglement of the shared entangled state.
\end{abstract}



\maketitle

\section{Introduction}
Quantum entanglement plays an essential role in various types of quantum
information processing. A notable example is the dense coding (sometimes
called superdense coding) originally proposed by Bennett and Wiesner \cite
{BW}. Its scheme is as follows. Suppose that the sender (Alice) and receiver
(Bob) initially share a maximally entangled pair of qubits [an
Einstein-Podolsky-Rosen (EPR) state], $\left| \Psi ^{-}\right\rangle =\left(
\left| \uparrow \right\rangle _{A}\left| \downarrow \right\rangle
_{B}-\left| \downarrow \right\rangle _{A}\left| \uparrow \right\rangle
_{B}\right) /\sqrt{2}$, where $\left| \uparrow \right\rangle
_{A(B)}=(1,0)^{t}$ and $\left| \downarrow \right\rangle _{A(B)}=(0,1)^{t}$.
Alice performs one of four possible unitary transformations $\{{\bf I}%
_{2},\sigma _{1},\sigma _{2},\sigma _{3}\}$ on her qubit, where ${\bf I}_{2}$
stands for the two-dimensional identity and $\sigma _{i}$ $(i=0,1,2,3)$ are
the Pauli matrices. According to her choice of transformations, the EPR
state is transformed into one of four mutually orthogonal states $\{\left|
\Psi ^{-}\right\rangle ,-\left| \Phi ^{-}\right\rangle ,\sqrt{-1}\left| \Phi
^{+}\right\rangle ,\left| \Psi ^{+}\right\rangle \}$, where $\left| \Psi
^{+}\right\rangle =\left( \left| \uparrow \right\rangle _{A}\left|
\downarrow \right\rangle _{B}+\left| \downarrow \right\rangle _{A}\left|
\uparrow \right\rangle _{B}\right) /\sqrt{2}$ and $\left| \Phi ^{\pm
}\right\rangle =\left( \left| \uparrow \right\rangle _{A}\left| \uparrow
\right\rangle _{B}\pm \left| \downarrow \right\rangle _{A}\left| \downarrow
\right\rangle _{B}\right) /\sqrt{2}$. Now, she sends off her qubit to Bob,
who performs an orthogonal measurement on the joint system of the received
qubit and his original one. The measured outcome unambiguously distinguishes
the signal state that Alice prepared. Thus, sending a{\it \ single} qubit
transmits $\log _{2}4=2$ bits of classical information. This is absolutely
impossible without entanglement; the amount of information conveyed by an 
{\it isolated} qubit cannot exceed one bit. Mattle \etal have
experimentally demonstrated dense coding transmission using
polarization-entangled photons \cite{MWKZ}. Barenco and Ekert \cite{BE} and
Hausladen \etal \cite{HJSWW} have argued about the generalization of
two-state systems in the Bennett-Wiesner dense coding scheme to $N$-state
quantum systems. Dense coding for continuous variables has also been
proposed by Braunstein and Kimble \cite{BK}. Bose, Plenio, and Vedral have
shown that the equal probabilities for signal states yield the maximum
capacity when the initially shared entangled states of two qubits are pure
states or Bell diagonal states under the condition that the set of unitary
transformations is restricted to $\{{\bf I}_{2},\sigma _{1},\sigma
_{2},\sigma _{3}\}$ \cite{BPV}. However, when the shared entangled state is
a general mixed one, the optimal dense coding scheme is still unknown. In
this paper, I prove that the dense coding scheme with the set of mutually
orthogonal unitary transformations and equal signal probabilities is optimal
for any entangled states in $C^{d}\otimes C^{d}$ shared between the sender
and receiver.

\section{Capacity for dense coding}
The general density matrix for a system on $C^{d}\otimes C^{d}$ is written
in the Hilbert-Schmidt representation as

\begin{equation}
\rho =\frac{1}{d^{2}}\left( {\bf I}_{d}\otimes {\bf I}_{d}+%
\sum_{i=1}^{d^{2}-1}r_{i}\lambda _{i}\otimes {\bf I}_{d}+{\bf I}_{d}\otimes
\sum_{i=1}^{d^{2}-1}s_{i}\lambda _{i}+\sum_{i,j=1}^{d^{2}-1}t_{ij}\lambda
_{i}\otimes \lambda _{j}\right) ,  \label{eq:Density}
\end{equation}
where $r_{i}$, $s_{i}$, and $t_{ij}$ are real numbers. In Eq.~(\ref
{eq:Density}) $\lambda _{i}$ $(i=1,2,\cdots ,d^{2}-1)$ are the generators of 
$SU(d)$ algebra satisfying

\begin{equation}
\Tr (\lambda _{i})=0.  \label{eq:Trace1}
\end{equation}
The generators $\lambda _{i}$ are given by \cite{SM}

\begin{equation}
\{\lambda _{i}\}_{i=1}^{d^{2}-1}=\{u_{1,2},u_{1,3},\cdots
,u_{d-1,d},v_{1,2},v_{1,3},\cdots ,v_{d-1,d},w_{1},w_{2},\cdots ,w_{d-1}\},
\label{eq:Generator}
\end{equation}
where

\begin{equation}
u_{i,j}=P_{i,j}+P_{j,i},  \label{eq:u}
\end{equation}
and

\begin{equation}
v_{i,j}=\sqrt{-1}(P_{i,j}-P_{j,i}),  \label{eq:v}
\end{equation}
with $1\leq i<j\leq d$, and

\begin{equation}
w_{k}=-\sqrt{\frac{2}{k(k+1)}}\left(
\sum_{i=1}^{k}P_{i,i}-kP_{k+1,k+1}\right) ,  \label{eq:w}
\end{equation}
with $1\leq k\leq d-1$. In Eqs.~(\ref{eq:u}), (\ref{eq:v}), and (\ref{eq:w}),

\begin{equation}
P_{i,j}=\left| i\right\rangle \left\langle j\right|  \label{eq:Projector}
\end{equation}
with $\{\left| i\right\rangle \}_{i=1}^{d}$ being
the orthonormal basis set on $C^{d}$; $\left| 1\right\rangle =(1,0,\cdots
,0)^{t},\left| 2\right\rangle =(0,1,,\cdots ,0)^{t},\cdots ,\left|
d\right\rangle =(0,0,\cdots ,1)^{t}$.

In general dense coding, Alice performs one of the local unitary
transformations $U_{i}\in U(d)$ on her $d$-dimensional quantum system to put
the initially shared entangled state $\rho $ in $\rho _{i}=(U_{i}\otimes 
{\bf I}_{d})\rho (U_{i}^{\dagger }\otimes {\bf I}_{d})$ with {\it a priori}
probability $p_{i}$ $(i=0,1,\cdots ,i_{max})$, and then she sends off her
quantum system to Bob. Upon receiving this quantum system, Bob performs a
suitable measurement on $\rho _{i}$ to extract the signal. The optimal
amount of information that can be conveyed is known to be bounded from above
by the Holevo quantity \cite{K},

\begin{equation}
\chi =S(\overline{\rho })-\sum_{i=0}^{i_{max}}p_{i}S(\rho _{i}),
\label{eq:Holevo}
\end{equation}
where $S(\rho )=-\Tr (\rho \log _{2}\rho )$ denotes the von Neumann
entropy and $\overline{\rho }=\sum_{i=0}^{i_{max}}p_{i}\rho _{i}$ is the
average density matrix of the signal ensemble. Since the Holevo quantity is
asymptotically achievable \cite{H,SW97}, I use Eq.~(\ref{eq:Holevo}) here as
the definition of the capacity of dense coding as in \cite{HJSWW,BPV}. Since
the von Neumann entropy is invariant under unitary transformations, $S(\rho
_{i})=S(\rho )$. Therefore, the dense coding capacity $\chi $ of Eq.~(\ref
{eq:Holevo}) can be rewritten as

\begin{equation}
\chi =S(\overline{\rho })-S(\rho ).
\end{equation}
It is also written as

\begin{equation}
\chi =\sum_{i=0}^{i_{max}}p_{i}S(\rho _{i}||\overline{\rho }),
\end{equation}
where $S(\rho ||\sigma )=\Tr \left[ \rho \left( \log _{2}\rho -\log
_{2}\sigma \right) \right] $ is the quantum relative entropy of $\rho $ with
respect to $\sigma $.

\section{Optimal capacity}
The problem is to find the optimal signal ensemble $\{\rho
_{i};p_{i}\}_{i=0}^{i_{max}}$ that maximizes $\chi $. Below I show that the
$d^{2}$ signal states ($i_{max}=d^{2}-1$) generated by mutually orthogonal unitary transformations with
equal probabilities yield the maximum $\chi $. This is the central result of
this paper. The mutually orthogonal unitary transformations are constructed
as

\begin{equation}
U_{i=(p,q)}\left| j\right\rangle =\exp \left( \sqrt{-1}\frac{2\pi }{d}%
pj\right) \left| j+q(%
\mathop{\rm mod}%
d)\right\rangle ,  \label{eq:Unitary}
\end{equation}
where integers $p$ and $q$ run from 0 to $d-1$ such that the number of
suffices $i$ is $d^{2}$; $0=(p=0,q=0),1=(p=0,q=1),\cdots
,d^{2}-1=(p=d-1,q=d-1)$. Note that $U_{i=0}={\bf I}_{d}$. The unitary
matrices thus defined satisfy the orthogonality relation, $d^{-1}\Tr%
\left( U_{i}^{\dagger }U_{j}\right) =\delta _{ij}$. From now on, the
ensemble of signal states generated by the unitary transformations of Eq.~(%
\ref{eq:Unitary}) with the equal probabilities $p_{i}=d^{-2}$ is denoted $%
{\cal E}^{*}$.

\begin{equation}
{\cal E}^{*}=\{(U_{i}\otimes {\bf I}_{d})\rho (U_{i}^{\dagger }\otimes {\bf I%
}_{d});p_{i}=d^{-2}\}_{i=0}^{d^{2}-1}.
\end{equation}
Furthermore, the capacity of dense coding with signal state ensemble ${\cal E%
}^{*}$ is denoted $\chi ^{*}$, which is given by $S(\overline{\rho }%
^{*})-S(\rho )$, where $\overline{\rho }^{*}=d^{-2}%
\sum_{i=0}^{d^{2}-1}(U_{i}\otimes {\bf I}_{d})\rho (U_{i}^{\dagger }\otimes 
{\bf I}_{d})$ is the average state of ${\cal E}^{*}$. In verifying the main
result (Theorem~\ref{thm:TH1}), the following three lemmas are crucial.

\begin{lem}
\label{lem:LM1}
The average state of ${\cal E}^{*}$ is separable and is given
by

\begin{equation}
\overline{\rho }^{*}=\frac{1}{d}{\bf I}_{d}\otimes \rho ^{B},
\label{eq:Average}
\end{equation}
where $\rho ^{B}=\Tr _{A}(\rho )$.
\end{lem}

\noindent{\bf Proof.} It is easy to show that

\begin{equation}
\sum_{i=0}^{d^{2}-1}U_{i}P_{j,k}U_{i}^{\dagger }=\delta _{jk}d{\bf I}_{d},
\label{eq:Matrix1}
\end{equation}
where $P_{j,k}$ is defined in Eq.~(\ref{eq:Projector}%
). Applying Eq.~(\ref{eq:Matrix1}) to the definition of $\lambda _{j}$ [Eq.~(%
\ref{eq:Generator}) with Eqs.~(\ref{eq:u}), (\ref{eq:v}), and (\ref{eq:w})],
we have

\begin{equation}
\sum_{i=0}^{d^{2}-1}U_{i}\lambda _{j}U_{i}^{\dagger }=0,  \label{eq:Matrix2}
\end{equation}
for $j=1,\cdots ,d^{2}-1$. Making use of Eq.~(\ref{eq:Matrix2}), $\overline{%
\rho }^{*}$ is calculated as

\begin{equation}
\overline{\rho }^{*}=\frac{1}{d^{2}}\sum_{i=0}^{d^{2}-1}(U_{i}\otimes {\bf I}%
_{d})\rho (U_{i}^{\dagger }\otimes {\bf I}_{d})=\frac{1}{d}{\bf I}%
_{d}\otimes \frac{1}{d}\left( {\bf I}_{d}+\sum_{i=1}^{d^{2}-1}s_{i}\lambda
_{i}\right) .
\end{equation}
This is clearly separable or disentangled. By noting that $\rho ^{B}=\Tr%
_{A}(\rho )=d^{-1}\left( {\bf I}_{d}+\sum_{i=1}^{d^{2}-1}s_{i}\lambda
_{i}\right) $, we readily obtain Eq.~(\ref{eq:Average}).  \hfill \fullsquare

\begin{lem}
\label{lem:LM2}
For any state $\omega $ written as ($U\otimes {\bf I}_{d})\rho
(U^{\dagger }\otimes {\bf I}_{d})$ with $U\in U(d)$, the quantum relative
entropy of $\ \omega $ with respect to $\overline{\rho }^{*}$ is equal to $%
\chi ^{*}$;

\begin{equation}
S(\omega ||\overline{\rho }^{*})=\chi ^{*}.  \label{eq:MDP}
\end{equation}
\end{lem}

\noindent{\bf Proof.} The density matrix $\rho $ of Eq.~(\ref{eq:Density}) is
rewritten as

\begin{equation}
\rho =\frac{1}{d}{\bf I}_{d}\otimes \rho ^{B}+\frac{1}{d^{2}}\left(
\sum_{i=1}^{d^{2}-1}r_{i}\lambda _{i}\otimes {\bf I}_{d}+%
\sum_{i,j=1}^{d^{2}-1}t_{ij}\lambda _{i}\otimes \lambda _{j}\right) .
\end{equation}
Therefore,

\begin{eqnarray}
\omega  &=&(U\otimes {\bf I}_{d})\rho (U^{\dagger }\otimes {\bf I}_{d}) 
\nonumber \\
&=&\frac{1}{d}{\bf I}_{d}\otimes \rho ^{B}+\frac{1}{d^{2}}\left[
\sum_{i=1}^{d^{2}-1}r_{i}(U\lambda _{i}U^{\dagger })\otimes {\bf I}%
_{d}+\sum_{i,j=1}^{d^{2}-1}t_{ij}(U\lambda _{i}U^{\dagger })\otimes \lambda
_{j}\right] \,.  \label{eq:Omega}
\end{eqnarray}
Now, from the result of Lemma~\ref{lem:LM1},

\begin{equation}
\log _{2}\overline{\rho }^{*}={\bf I}_{d}\otimes \log _{2}\left( \frac{\rho
^{B}}{d}\right) .  \label{eq:Log}
\end{equation}
From Eqs.~(\ref{eq:Omega}) and (\ref{eq:Log}), 
\begin{eqnarray}
&&\Tr (\omega \log _{2}\overline{\rho }^{*})=\Tr (\overline{\rho }%
^{*}\log _{2}\overline{\rho }^{*})  \nonumber \\
&&\qquad \qquad \qquad +\frac{1}{d^{2}}\left\{ \sum_{i=1}^{d^{2}-1}r_{i}\Tr \left[ (U\lambda _{i}U^{\dagger })\otimes \log _{2}\left( \frac{\rho ^{B}%
}{d}\right) \right] \right.  \nonumber\\
&&\left. \qquad \qquad \qquad \qquad +\sum_{i,j=1}^{d^{2}-1}t_{ij}\Tr%
\left[ (U\lambda _{i}U^{\dagger })\otimes \lambda _{j}\log _{2}\left( \frac{%
\rho ^{B}}{d}\right) \right] \right\} .  \label{eq:Trace2}
\end{eqnarray}
By using the formula $\Tr (A\otimes B)=\Tr (A) \Tr (B)$ and the
properties of $\lambda _{i}$ of Eq.~(\ref{eq:Trace1}), the last term of the
right-hand side of Eq.~(\ref{eq:Trace2}) vanishes; $\Tr (\omega \log _{2}%
\overline{\rho }^{*})=\Tr (\overline{\rho }^{*}\log _{2}\overline{\rho }%
^{*})=-S(\overline{\rho }^{*})$. We thus obtain

\begin{eqnarray}
S(\omega ||\overline{\rho }^{*}) &=&\Tr \left[ \omega \left( \log
_{2}\omega -\log _{2}\overline{\rho }^{*}\right) \right] \nonumber \\
&=&-S(\omega )+S(\overline{\rho }^{*})=-S(\rho )+S(\overline{\rho }^{*}).
\label{eq:Relative}
\end{eqnarray}
In the last line of Eq.~(\ref{eq:Relative}), the equality $S(\omega )=S(\rho
)$ was used. Since $\chi ^{*}=S(\overline{\rho }^{*})-S(\rho )$, $S(\omega ||%
\overline{\rho }^{*})=\chi ^{*}$. This completes the proof.  \hfill \fullsquare

\begin{lem}
\label{lem:LM3}
The average quantum relative entropy of signal ensemble $\{\rho
_{k};p_{k}\}$ with respect to a density matrix $\rho ^{\prime }$ is given by
\begin{equation}
\sum_{k}p_{k}S(\rho _{k}||\rho ^{\prime })=\sum_{k}p_{k}S(\rho _{k}||%
\overline{\rho })+S(\overline{\rho }||\rho ^{\prime }),
\label{eq:Donald}
\end{equation}
where $p_{k}\geq 0$, $\sum_{k}p_{k}=1$, and $\overline{\rho }%
=\sum_{k}p_{k}\rho _{k}$.
\end{lem}

Equation (\ref{eq:Donald}) is known as Donald's identity \cite{D}.

\begin{thm}
\label{thm:TH1}
The dense coding capacity $\chi ^{*}$ is maximum. That is,
for all possible signal ensembles $\{\omega _{i};q_{i}\}_{i=0}^{i_{max}}$,

\begin{equation}
\chi ^{*}\geq \sum_{i=0}^{i_{max}}q_{i}S(\omega _{i}||\overline{\omega }),
\end{equation}
where $\overline{\omega }=\sum_{i=0}^{i_{max}}q_{i}\omega _{i}$.
\end{thm}

\noindent{\bf Proof.} Since $S(\omega _{i}||\overline{\rho }^{*})=\chi ^{*}$ for $%
i=0,1,\cdots ,i_{max}$ (Lemma~\ref{lem:LM2}),

\begin{equation}
\chi ^{*}=\sum_{i=0}^{i_{max}}q_{i}S(\omega _{i}||\overline{\rho }^{*}).
\label{eq:Optimal}
\end{equation}
Applying Donald's identity of Lemma~\ref{lem:LM3} [Eq.~(\ref{eq:Donald})] to the
right-hand side of Eq.~(\ref{eq:Optimal}), we obtain $\chi ^{*}=\chi +S(%
\overline{\omega }||\overline{\rho }^{*})$, where $\chi
=\sum_{i=0}^{i_{max}}q_{i}S(\omega _{i}||\overline{\omega })$, the dense
coding capacity with ensemble $\{\omega _{i};q_{i}\}_{i=0}^{i_{max}}$. Since
the relative entropy is strictly non-negative; $S(\overline{\omega }||%
\overline{\rho }^{*})\geq 0$, $\chi ^{*}\geq \chi $. That is, $\chi ^{*}$ is
indeed the optimal dense coding capacity; i.e., ${\cal E}^{*}$ is the
optimal signal ensemble. This completes the proof.  \hfill \fullsquare

Equation~(\ref{eq:MDP}) means that the average ensemble $\overline{\rho }%
^{*} $ has the {\it maximal distance property} \cite{SW01}; that is, $S(\omega
||\overline{\rho }^{*})$ cannot exceed $\chi ^{*}$ for any $\omega
=(U\otimes {\bf I}_{d})\rho (U^{\dagger }\otimes {\bf I}_{d})$. Theorem~\ref{thm:TH1} is
also the direct consequence of this fact. Note that the optimal dense coding
scheme for $d=2$ is reduced to Bennett and Wiesner's scheme.

\section{Bounds on optimal capacity}
Next, I prove the following theorem concerning the bounds on $\chi ^{*}$.

\begin{thm}
\label{thm:TH2}
The optimal capacity $\chi ^{*}$ satisfies

\begin{equation}
E_{R}(\rho )\leq \chi ^{*}\leq E_{R}(\rho )+\log _{2}d,  \label{eq:Bounds}
\end{equation}
where $E_{R}(\rho )$ is the relative entropy of entanglement of $\rho $.
\end{thm}

The relative entropy of entanglement is defined as $E_{R}(\rho
)=\min_{\sigma \in {\cal D}}S(\rho ||\sigma )$, where the minimum is taken
over ${\cal D}$, the set of all disentangled states \cite{VP}. The proof of
the first inequality of Eq.~(\ref{eq:Bounds}) is essentially the same as
that given in \cite{BPV} for $d=2$. By noting that $\overline{\rho }^{*}$ is
a disentangled state (Lemma~\ref{lem:LM1}), we get

\begin{equation}
S(\rho _{i}||\overline{\rho }^{*})\geq \min_{\sigma \in {\cal D}}S(\rho
_{i}||\sigma )=E_{R}(\rho _{i}).
\end{equation}
Consequently,

\begin{equation}
\chi ^{*}=\frac{1}{d^{2}}\sum_{i=0}^{d^{2}-1}S(\rho _{i}||\overline{\rho }%
^{*})\geq \frac{1}{d^{2}}\sum_{i=0}^{d^{2}-1}E_{R}(\rho _{i}).
\end{equation}
Since the relative entropy of entanglement is invariant under local unitary
operations \cite{VP}, $E_{R}(\rho _{i})=E_{R}\left[ (U_{i}\otimes {\bf I}%
_{d})\rho (U_{i}^{\dagger }\otimes {\bf I}_{d})\right] =E_{R}(\rho )$.
Therefore, $\chi ^{*}\geq E_{R}(\rho )$. The second part of the inequality
in (\ref{eq:Bounds}) for $d=2$ has been conjectured previously in \cite{BPV}%
. In the proof of this inequality, the following relation given by Plenio,
Virmani, and Papadopoulos \cite{PVP},

\begin{equation}
\max \{S(\rho ^{A})-S(\rho ),S(\rho ^{B})-S(\rho )\}\leq E_{R}(\rho ),
\end{equation}
plays a key role. It implies that

\begin{equation}
S(\rho ^{B})-S(\rho )\leq E_{R}(\rho ).  \label{eq:PVP}
\end{equation}
Now, from Eqs.~(\ref{eq:Average}) and (\ref{eq:Log}), we have

\begin{eqnarray}
S(\overline{\rho }^{*}) &=&-\Tr (\overline{\rho }^{*}\log _{2}\overline{%
\rho }^{*})  \nonumber \\
&=&-\Tr \left[ \left( {\bf I}_{d}\otimes \frac{\rho ^{B}}{d}\right)
\left( {\bf I}_{d}\otimes \log _{2}\frac{\rho ^{B}}{d}\right) \right]  
\nonumber \\
&=&-\Tr ({\bf I}_{d})\Tr \left( \frac{\rho ^{B}}{d}\log _{2}\frac{%
\rho ^{B}}{d}\right) =S(\rho ^{B})+\log _{2}d.  \label{eq:Entropy}
\end{eqnarray}
In the last line of Eq.~(\ref{eq:Entropy}), the fact that $\Tr (\rho
^{B})=1$ was used. Substituting Eq.~(\ref{eq:Entropy}) into the left-hand
side of (\ref{eq:PVP}), we readily obtain

\begin{equation}
S(\overline{\rho }^{*})-S(\rho )\leq E_{R}(\rho )+\log _{2}d.
\label{eq:Bound2}
\end{equation}
Since the left-hand side of (\ref{eq:Bound2}) is just $\chi ^{*}$, we have $%
\chi ^{*}\leq E_{R}(\rho )+\log _{2}d$. For $d=2$, it has been proved that
the equality holds when $\rho $ is the Bell diagonal state with only two
non-zero eigenvalues \cite{BPV}.

\section{Conclusions}
In summary, it has been proved that optimal dense coding with a general
entangled state on the Hilbert space $C^{d}\otimes C^{d}$ is achieved when
the sender prepares the signal states by mutually orthogonal unitary
transformations with equal {\it a priori} probabilities. It is also proved
that the optimal capacity of dense coding $\chi ^{*}$ satisfies $E_{R}(\rho
)\leq \chi ^{*}\leq E_{R}(\rho )+\log _{2}d$, where $E_{R}(\rho )$ is the
relative entropy of entanglement of the shared entangled state.


\section*{References}

\begin{thebibliography}{99}

\bibitem{BW} Bennett C H and  Wiesner S J 1992 \PRL {\bf 69} 2881

\bibitem{MWKZ} Mattle K, Weinfurter H, Kwiat P G and Zeilinger A 1996 \PRL {\bf 76} 4656

\bibitem{BE}  Barenco A and Ekert A 1995 {\it J. Mod. Opt.} {\bf 42} 1253

\bibitem{HJSWW}  Hausladen P, Jozsa R, Schumacher B, Westmoreland M and
Wootters W K 1996 \PR A {\bf 54} 1869

\bibitem{BK}  Braunstein S L and Kimble H J 2000 \PR A {\bf 61} 042302-1

\bibitem{BPV}  Bose S, Plenio M B and Vedral V 2000 {\it J. Mod. Opt.} {\bf 47} 291

\bibitem{SM}  Schlienz J and Mahler G 1995 \PR A {\bf 52} 4396

\bibitem{K}  Kholevo A S 1973 {\it Probl. Peredachi Inf.} {\bf 9} 3 [1973 {\it Probl.
Inf. Transm. (USSR)} {\bf 9} 110]

\bibitem{H}  Holevo A S 1998 {\it IEEE Trans. Inf. Theory} {\bf 44} 269

\bibitem{SW97}  Schumacher B and Westmoreland M D 1997 \PR A {\bf 56} 131

\bibitem{D}  Donald M J 1987 {\it Math. Proc. Cam. Phil. Soc.} {\bf 101} 363

\bibitem{VP}  Vedral V and Plenio M B 1998 \PR A {\bf 57} 1619

\bibitem{SW01}  Schumacher B and Westmoreland M D 2001 \PR A {\bf 63} 022308-1; {\it Preprint} quant-ph/0004045

\bibitem{PVP}  Plenio M B, Virmani S and Papadopoulos P 2000 \JPA {\bf 33} L193

\end{thebibliography}
\end{document}